\documentclass[aps,prd,onecolumn,amsmath,10pt,superscriptaddress,floatfix,nofootinbib]{revtex4-1}

\usepackage{epsfig,amssymb,amsfonts,amsmath,mathtools,bm,color,xcolor,graphicx,braket,adjustbox,esint,upgreek,subfigure}
\usepackage{multirow}
\usepackage[utf8]{inputenc}

\usepackage{amsmath} 
\usepackage{amssymb} 
\usepackage{amsfonts} 
\usepackage{mathrsfs}
\usepackage{slashed} 
\usepackage{graphicx}
\usepackage{subfigure}
\usepackage{caption}
\usepackage{booktabs}
\usepackage{threeparttable}
\usepackage{multirow}
\usepackage{color}
\usepackage{verbatimbox}
\usepackage{appendix}
\newcommand{\be}{\begin{equation}}
\newcommand{\ee}{\end{equation}}
\newcommand{\bea}{\begin{eqnarray}}
\newcommand{\eea}{\end{eqnarray}}
\newcommand{\beas}{\begin{eqnarray*}}
	\newcommand{\eeas}{\end{eqnarray*}}

\newcommand{\lzu}{\affiliation{School of Physical Sciences, Lanzhou University, Lanzhou 730000, China}}

\newcommand{\imp}{\affiliation{Institute of Modern Physics,\\ Chinese Academy of Sciences, Lanzhou 730000, China}}

\newcommand{\ucas}{\affiliation{School of Physical Sciences, University of Chinese Academy of Sciences, Beijing 100049, China}}

\newcommand{\scnu}{\affiliation{Guangdong Provincial Key Laboratory of Nuclear Science, Institute of Quantum Matter, South China Normal University, Guangzhou 510006, China}}

\graphicspath{{FiguresV2/}}

\begin{document}
	
\title{Solitary wave solutions of FKPP equation using Homogeneous balance method(HB method)}

\author{Yirui Yang }
\lzu
\author{Wei Kou }
\imp \ucas
\author{Xiaopeng Wang }
\imp \ucas
\author{Xurong Chen}\email{xchen@impcas.ac.cn}
\imp \ucas \scnu

\begin{abstract}
	In this paper, we use the homogeneous balance(HB) method is used to construct function transformation to solve the nonlinear development equation——Fisher-Kolomogror-Pertrovskii-Piskmov equation (FKPP equation), then the exact solution of FKPP equation is obtained, and the solution is converted into the form of isolated wave solution. Finally, we have shown the solution of FKPP equation in some picture forms, and the rationality of the solution in this paper can be verified by using the pictures.

\end{abstract}

\maketitle

%\vspace{1cm}

\section{Introduction}

Over the last decades, the nonlinear science has rapidly developed into the frontier field of modern science and technology research. For the nonlinear science, the solution of nonlinear equations has always been the difficult and the hot argument. One way to solve the nonlinear equation is to construct a transformation of a nonlinear function to solve it. For some one-dimensional nonlinear equations, the HB method is used to construct function transformation to solve them tentatively.

On the other hand, soliton theory is a very important branch of nonlinear science, and the research of soliton theory is also a hot topic at present. It plays an increasingly extensive role in many fields of mathematics and physics. The soliton theory is of great practical value, because it is widely used in fluid mechanics, nonlinear optics and other fields. Furthermore, some physical phenomenons that cannot be explained by classical theory but can be explained by using the soliton theory. 

Fisher-Kolomogror-Pertrovskii-Piskmov equation (FKPP equation) is a basic models of mathematical biology and ecology, this equation is a typical convection-reaction-diffusion equation. Because it involves a large number of problems from physics, chemistry, biology, population dynamics and other disciplines, this equation has a very broad practical background.

In this paper, we try to use HB method to obtain the exact solution of FKPP equation, and the solution is reduced to the form of the solitary wave solutions.

\section{The solution of FKPP equation}

 FKPP equation can be written as
\begin{equation}
u_t-u_{xx}-u+u^2=0.
\end{equation}
according to the HB method in ref.[1-4], we assume that the solution of FKPP equation canbe expanded by the form
\begin{equation}
u(x, t)=\sum_{m+n=1}^{N} a_{m+n} \frac{\partial^{(m+n)}}{\partial x^{m} \partial t^{n}} f(\varphi(x, t)),
\end{equation}
where $m$ and $n$ are non-negative integers, the power term of the highest-order partial derivative of $\varphi \left( x,t \right)$ appearing in the formula is $f^{\left( N \right)}\left( \varphi \right) \varphi _x^m\varphi _t^n$, with $m+n=N$. $f\left( \varphi \right)$ and $\varphi \left( x,t \right)$ are functions to be solved by HB principle, and $N$ will be determined through the HB principle.

Set the indicated number of highest-order derivatives as $N=m+n$ in the form of solution $u\left( x,t \right)$, take $a_{m+n}=1$ then the highest-order partial derivative term becomes $f^{\left( N \right) }\left( \varphi \right) \varphi _x^m\varphi _t^n$. Substitute it into eq.(1), then the non-linear term $u^2$ is written by
\begin{equation}
u^2=f^{\left( N \right)}f^{\left( N \right)}\varphi _x^{2m}\varphi _t^{2n}+\cdots,
\end{equation}
the highest-order partial derivative term in the equation reads
\begin{equation}
u_{xx}=f^{\left( N+2 \right)}f^{\left( N \right)}\varphi _x^{\left( m+2 \right)}\varphi _t^n+\cdots.
\end{equation}

By the principle of partial balance, the power of $\varphi _x$ and $\varphi _t$ should be balanced, so we get
\begin{equation}
\begin{aligned}
2m=m+2,\ \ \ \ \ \ 2n=n,
\end{aligned}
\end{equation}
then we obtain $m=2,\ n=0$, then $N=m+n=2$ can be known, the solution of FKPP equation can be rewritten like the form
\begin{equation}
u\left( x,t \right) =f_{xx}+af_x+b=f''\left( \varphi \right) \varphi _x^2+f'\varphi _xx+af'\varphi _x+b,
\end{equation}
substitution of eq. (6) in every term of (1) yields, then we have
\begin{equation}
\begin{aligned}
u_t&=f^{\left( 3 \right)}\varphi _{x}^{2}\varphi _t+f''\left( 2\varphi _x\varphi _{xt}+\varphi _{xx}\varphi _t+a\varphi _x\varphi _t \right) +f'\left( \varphi _{xxt}+a\varphi _{xt} \right),\\
u_{xx}&=f^{\left( 4 \right)}\varphi _{x}^{4}+f^{\left( 3 \right)}\left( 6\varphi _{x}^{2}\varphi _{xx}+a\varphi _{x}^{3} \right)+f''\left( 3\varphi _{xx}^{2}+4\varphi _x\varphi _{xxx}+3a\varphi _x\varphi _{xx} \right) +f'\left( \varphi _{xxxx}+a\varphi _{xxxx} \right),\\
u&=f''\varphi _x^2+f'\varphi _{xx}+af'\varphi _x+b,\\
u^2&=f''^2\varphi _{x}^{2}+f'f''\left( 2\varphi _{x}^{2}\varphi _{xx}+2a\varphi _{x}^{3} \right) +2bf''\varphi _{x}^{2}+f'^2\left( \varphi _{xx}^{2}+a^2\varphi _{x}^{2}+2a\varphi _{xx}\varphi _x \right) +f'\left( 2b\varphi _{xx}+2ab\varphi _x \right) +b^2.
\end{aligned}
\end{equation}

Observing the terms on the right of eq. (6) and (7) above, and identify the highest power of the $\varphi _x$ term which is $\varphi _x^4$, merge to get
\begin{equation}
(f''^2-f^{\left( 4 \right)})\varphi _x^4.
\end{equation}

Using the balance principle once again, the coefficient must be zero, then we can get the non-linear ODE which is dependent by $f$:
\begin{equation}
f''^2-f^{\left( 4 \right)}=0.
\end{equation}

One particular solution to this equation is
\begin{equation}
f=-6\ln\varphi,
\end{equation}
if we observe the non-linear terms on the right hand side of (7), which contain the derivative of $f,f''^2,f'f'',f'^2$, substitute the solution $f$ into eq.(9), we obtain
\begin{equation}
\begin{aligned}
f''^2&=f^{\left( 4 \right)},\\
f'f''&=3f^{\left( 3 \right)},\\
f'^2&=6f''.
\end{aligned}
\end{equation}

Considering all of these relations into eq.(7) and subtracting the term $\varphi _x^4$ satisfying eq. (11), the FKPP equation then becomes
\begin{equation}
\begin{aligned}
f^{\left( 3 \right)}\left( \varphi _{x}^{2}\varphi _t+5a\varphi _{x}^{2} \right)&+f''\left[ 2\varphi _x\varphi _{xt}+\varphi _{xx}\varphi _t+a\varphi _x\varphi _t+3\varphi _{xx}^{2}-4\varphi _x\varphi _{xxx}+9a\varphi _x\varphi _xx+\left( 2b+6a^2-1 \right) \varphi _{x}^{2} \right]\\ &+f''\left[ \varphi _{xxt}+a\varphi _{xt}-\varphi _{xxxx}-a\varphi _{xxx}+\left( 2b-1 \right) \varphi _{xx}+\left( 2ab-a \right) \varphi _x \right] +\left( b^2-b \right) =0.
\end{aligned}
\end{equation}

In the above equation, the coefficients of all terms can be set to zero. First, the constant terms $b^2-b=0$, then we get
\begin{equation}
b=0,\ \ b=1.
\end{equation}

Due to $\varphi \left( x,t \right)$, we can make an ansatz solution
\begin{equation}
\varphi \left( x,t \right) =1+e^{\left( kx+ct \right)}.
\end{equation}
Substituting it into eq. (11) , combining with the three equations and finding the undetermined coefficients $a,k,c$, that is
\begin{equation}
\begin{aligned}
&c+5ak=0,\\
&k^3-9ak^2-\left( 3c+2b+6a^2-1 \right) k-ac=0,\\
&k^3-ak^2-\left( 2b+c-1 \right) k-\left( 2ab+ac-a \right) =0,
\end{aligned}
\end{equation}
the equations are not independent and they can be reduced to
\begin{equation}
\begin{aligned}
&k=\left( -c \right) /5a,\\
&10ak^2+\left( 6a^2+2c \right) k+\left( -2ab+a \right) =0.
\end{aligned}
\end{equation}
Substitute (14) and (10) into eq. (6), and the general solution of the equation is obtained
\begin{equation}
u\left( x,t \right) =\frac{6e^{2\left( kx+ct \right)}k^2}{\left( 1+e^{kx+ct} \right) ^2}-\frac{6e^{kx+ct}k^2}{1+e^{kx+ct}}-\frac{6ae^{kx+ct}k}{1+e^{kx+ct}}+b.
\end{equation}

\section{Results and Discussion}

In order to solve this problem in the isolated waves form, we made the following analysis: According to the whole solution process, we can see that the form of the final solution depends on the eq.(14), (10) and (6).

Eq.(14) is a heuristic solution. In the case of the indicated number of highest-order derivatives $N=m+n=2$, we have eq.(5). Eq.(10) is the solution of the nonlinear ODE we came up with. According to the difference of the equation, the solutions we got have the following form
\begin{equation}
f=-m\ln\varphi,
\end{equation}
just with m=6.

According to the eq.(14), (10) and (6), $u(x,t)$ can be wriiten as
\begin{equation}
\begin{aligned}
u\left( x,t \right)&=f_{xx}+af_x+b\\
&=f''\left( \varphi \right) \varphi _{x}^{2}+f'\varphi _xx+af'\varphi _x+b\\
&=\left( -mk^2 \right) \left[ \frac{\varphi e^{kx+ct}}{\varphi ^2}-\frac{e^{2\left( kx+ct \right)}}{\varphi ^2} \right] +\left( -amk \right) \frac{e^{kx+ct}}{\varphi}+b,
\end{aligned}
\end{equation}
where
\begin{equation}
\begin{aligned}
\frac{\varphi e^{kx+ct}}{\varphi ^2}-\frac{e^{2\left( kx+ct \right)}}{\varphi ^2}&=\frac{\left( 1+e^{kx+ct} \right) e^{kx+ct}-e^{2\left( kx+ct \right)}}{\left( 1+e^{kx+ct} \right) ^2}\\
&=\frac{e^{kx+ct}}{\left( 1+e^{kx+ct} \right) ^2},
\end{aligned}
\end{equation}
and we can intrduce the Hyperbolic Secant function
\begin{equation}
\begin{aligned}
&\mathrm{sech}\ x=\frac{2}{e^x+e^{-x}},\\
&\mathrm{sech}^2x=\frac{4}{\left( e^x+e^{-x} \right) ^2}=\frac{4e^2x}{\left[ e^x\left( e^x+e^{-x} \right) \right] ^2}=\frac{4e^2x}{\left( e^2x+1 \right) ^2},\\
&\mathrm{sech}^2\left[ \frac{1}{2}\left( kx+ct \right) \right]=\frac{4e^{kx+ct}}{\left( e^{kx+ct}+1 \right) ^2}.
\end{aligned}
\end{equation}

We see eq.(20) and obtain
\begin{equation}
\begin{aligned}
\frac{\varphi e^{kx+ct}}{\varphi ^2}-\frac{e^{2\left( kx+ct \right)}}{\varphi ^2}=\frac{e^{kx+ct}}{\left( 1+e^{kx+ct} \right) ^2}=\frac{1}{4}\mathrm{sech}^2\left[ \frac{1}{2}\left( kx+ct \right) \right].
\end{aligned}
\end{equation}

Now let's analyze the second term of the eq.(19) with intrducing the Hyperbolic Tangent function, we have
\begin{equation}
\begin{aligned}
&\tanh x=\frac{e^x-e^{-x}}{e^x+e^{-x}}=\frac{e^x\left( e^x-e^{-x} \right)}{e^x\left( e^x+e^{-x} \right)}=\frac{e^{2x}-1}{e^{2x}+1},\\
&1+\tanh x=1+\frac{e^x-e^{-x}}{e^x+e^{-x}}=\frac{e^{2x}-1+e^{2x}+1}{e^{2x}+1}=\frac{2e^{2x}}{e^{2x}+1},\\
&1+\tanh\left[ \frac{1}{2}\left( kx+ct \right) \right] =\frac{2e^{kx+ct}}{e^{kx+ct}+1},\\
&\frac{e^{kx+ct}}{\varphi}=\frac{e^{kx+ct}}{e^{kx+ct}+1}=\frac{1}{2}\left\{ 1+\tanh\left[ \frac{1}{2}\left( kx+ct \right) \right] \right\}.
\end{aligned}
\end{equation}
Further, we have
\begin{equation}
\begin{aligned}
u\left( x,t \right)&=f_{xx}+af_x+b\\
&=f''\left( \varphi \right) \varphi _x^2+f'\varphi _{xx}+af'\varphi _x+b\\
&=\frac{\left( -mk^2 \right)}{4}\mathrm{sech}^2\left[ \frac{1}{2}\left( kx+ct \right) \right] +\frac{\left( -amk \right)}{2}\tanh\left[ \frac{1}{2}\left( kx+ct \right) \right] +\frac{\left( -amk \right)}{2}+b.
\end{aligned}
\end{equation}

From here we can see that the form of a general solution of $u(x,t)$ is only related to the eq.(18),(14) and (6). When the indicated number of highest-order derivatives reads $N=2$, and the ansatz solution is $\varphi \left( x,t \right) =1+e^{\left( kx+ct \right)}$, we also have such general solution.

To determine the exact solution, we just need to determine the parameters $m,a,k,c$ and $b$.
the values of these parameter shall be determined by the (6). Let the highest power coefficient be zero, i.e. $m = 6$, then we can get
\begin{equation}
\begin{aligned}
k&=\frac{2ab-a}{6a^2}=\frac{2b-1}{6a},\\
c&=-5a\frac{2b-1}{6a}=-\frac{5}{6}\left( 2b-1 \right).
\end{aligned}
\end{equation}

By substituting the obtained $k$ and $c$ into eq.(14) and the eq.(10) into the eq.(6), two accurate solitary wave solutions of FKPP equation can be obtained
\begin{equation}
u\left( x,t \right) =-\frac{3}{2}k^2\mathrm{sech}^2\left[ \frac{1}{2}\left( kx+ct \right) \right] +3ak\tanh\left[ \frac{1}{2}\left( kx+ct \right) \right] -3ak+b.
\end{equation}

Let us consider the following two cases,

\ \ (i) $b=0,c=\frac{5}{6},k=-\frac{1}{6a}$, the form of the solutions is
\begin{equation}
u\left( x,t \right) =-\frac{1}{24a^2}\mathrm{sech}^2\left[ \frac{1}{2}\left( -\frac{1}{6a}x+\frac{5}{6}t \right) \right] +\frac{1}{2}\tanh\left[ \frac{1}{2}\left( -\frac{1}{6a}x+\frac{5}{6}t \right) \right] +\frac{1}{2}
\end{equation}

\ \ (ii) $b=1,c=-\frac{5}{6},k=\frac{1}{6a}$, the form of the solutions is
\begin{equation}
u\left( x,t \right) =-\frac{1}{24a^2}\mathrm{sech}^2\left[ \frac{1}{2}\left( \frac{1}{6a}x-\frac{5}{6}t \right) \right] -\frac{1}{2}\tanh\left[ \frac{1}{2}\left( \frac{1}{6a}x-\frac{5}{6}t \right) \right] +\frac{1}{2}
\end{equation}

Although we have an exact solution with the parameter $a$, it is worth thinking about what the value of $a$ is. Since $a$ is a parameter that we set, obviously $a$ cannot be 0. We can substitute eq.(27) into FKPP equation to determine the value of $a$. It's very easy to get that $a$ can only be $\pm1/\sqrt{6}$.
Finally, we get the following figures as FIG. 1 are discussed for the solutions.

\begin{figure}
	\begin{center}
		\subfigure[ ]{\label{figa}
			\includegraphics[width=0.45\textwidth]{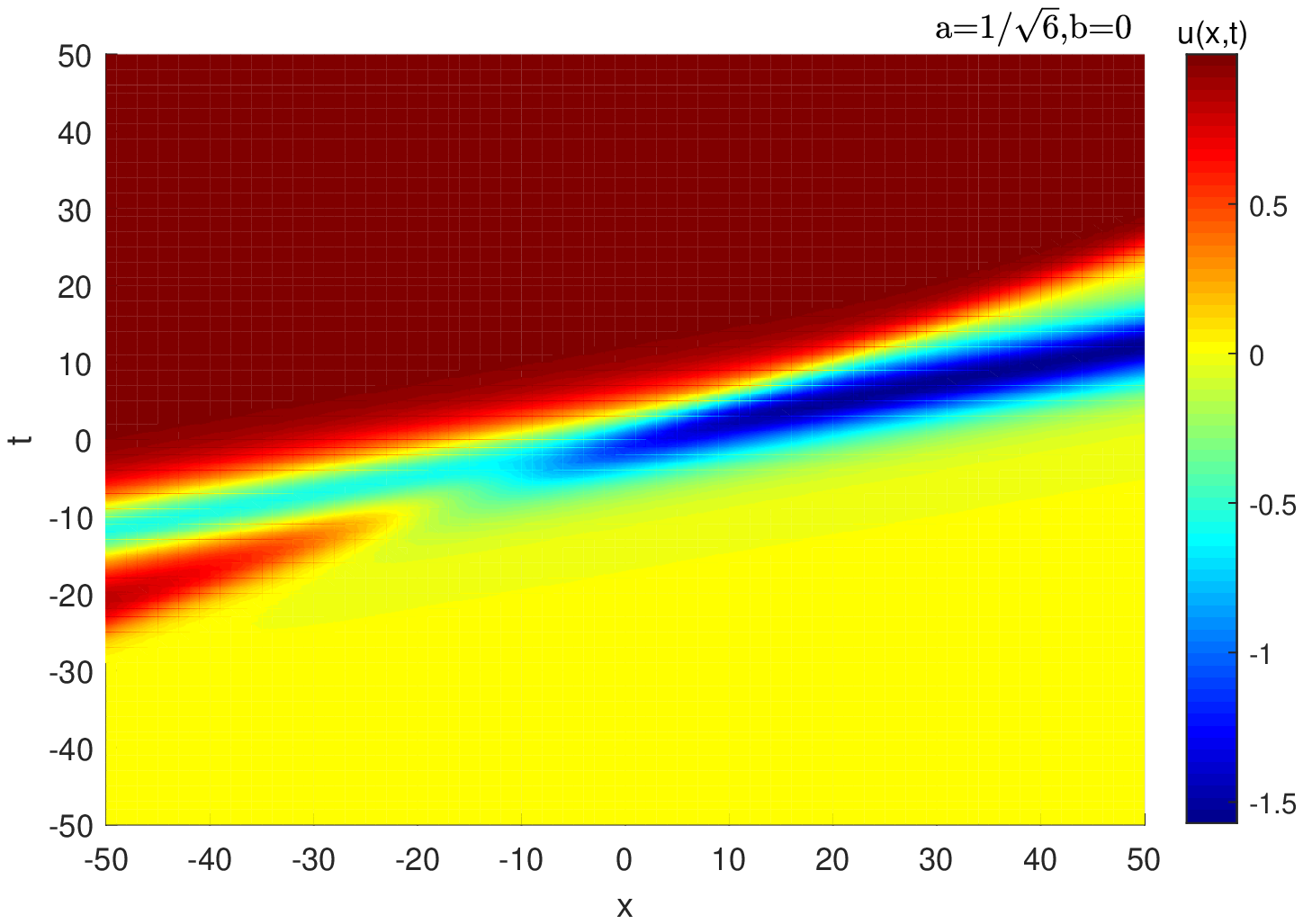}}
		\subfigure[ ]{\label{figb}
			\includegraphics[width=0.45\textwidth]{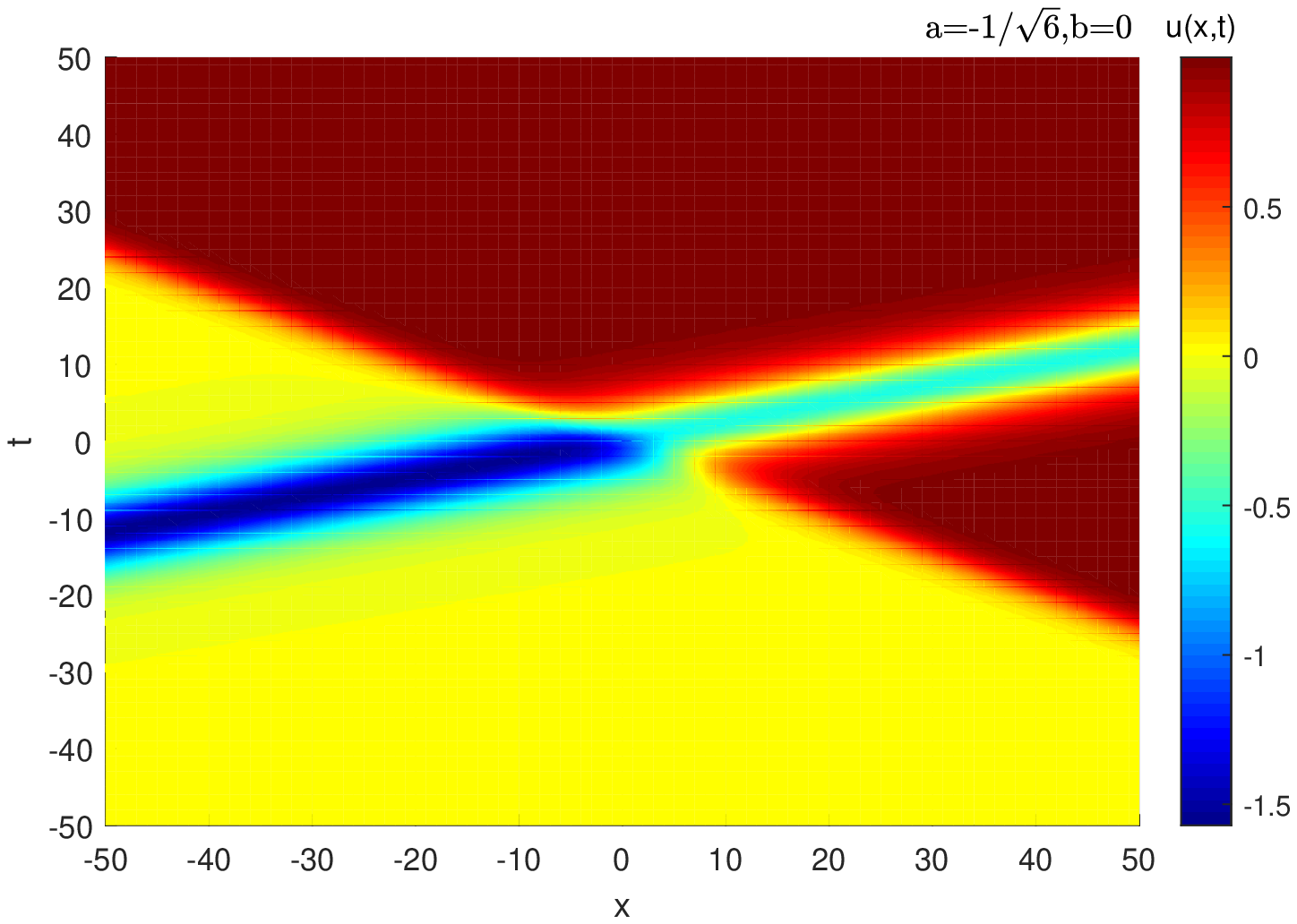}}
	\end{center}
	\caption{(a) $u(x,t)$ with $a=1/\sqrt{6},\ b=0$. (b) $u(x,t)$ with $a=-1/\sqrt{6},\ b=0$.
	}
	\label{fig2}
\end{figure}

\section{Summary}

In this paper, the FKPP equation is solved mainly by the  HB method, and the structural form of eq.(2) is mainly adopted for the transformation of the nonlinear function of the constructive solution. On this basis, the function transformation form of various solutions can be extended. Further more, the various solutions of the kind of equation can be obtained. The principle of the HB method is often used in the methods of hyperbolic function expansion and elliptic function expansion of nonlinear equations. It is worthy to discuss in the future.

\hspace{1.0cm}
\begin{acknowledgements}
We are grateful to K. Q. Yang for useful comments and discussions.
This work is supported by the Strategic Priority Research Program of Chinese Academy of Sciences, Grant No. XDB34030301.
\end{acknowledgements}


\begin{thebibliography}{99}
\bibitem{Wang1996} M. Wang, Phys. Lett. A 199, 169 (1995); 213, 279 (1996).
\bibitem{Wang2003} Y. B. Zhou, M. L. Wang, and Y. M. Wang, Phys. Lett. A 308,31 (2003).
\bibitem{Wang2004} Y. B. Zhou, M. L. Wang, and T. D. Miao, Phys. Lett. A 323, 77 (2004).
\bibitem{Wang2014} Wang, M. and Li, X. (2014), Journal of Applied Mathematics and Physics, 2, 823-827. doi: 10.4236/jamp.2014.28091.


\end{thebibliography}
\end{document}